\documentclass{webofc}
\usepackage[varg]{txfonts}   
\usepackage{color}
\definecolor{purple}{rgb}{0.5,0,0.5}
\definecolor{blue}{rgb}{0.0,0,0.9}
\definecolor{prdblue}{rgb}{0.133,0.118,0.498}
\usepackage[colorlinks=true, pdfstartview=FitV, linkcolor=prdblue, citecolor= prdblue, urlcolor=prdblue]{hyperref}

\usepackage[mathscr,scaled=1.15]{urwchancal}
\DeclareFontFamily{OT1}{pzc}{}
\DeclareFontShape{OT1}{pzc}{m}{it}%
{<-> s * [1.15] pzcmi7t}{}
\DeclareMathAlphabet{\mathpzc}{OT1}{pzc}{m}{it}

\begin{document}
\title{$\,$\\[-6ex]\hspace*{\fill}{\normalsize{\sf\emph{Preprint no}.\ NJU-INP 067/22}}\\[1.25ex]
Origin of the Proton Mass}
%
%

\author{\firstname{} \lastname{Craig D. Roberts}\inst{1,2}\fnsep\thanks{\email{cdroberts@nju.edu.cn}}
}

\institute{School of Physics, Nanjing University, Nanjing, Jiangsu 210093, China
\and
           Institute for Nonperturbative Physics, Nanjing University, Nanjing, Jiangsu 210093, China
          }

\abstract{%
  Atomic nuclei lie at the core of everything visible; and at the first level of approximation, their atomic weights are simply the sum of the masses of all the neutrons and protons (nucleons) they contain.  Each nucleon has a mass $m_N \approx 1\,{\rm GeV}\approx 2000$-times the electron mass.  The Higgs boson -- discovered at the large hadron collider in 2012, a decade ago -- produces the latter, but what generates the nucleon mass?  This is a pivotal question.  The answer is widely supposed to lie within quantum chromodynamics (QCD), the strong-interaction piece of the Standard Model.  Yet, it is far from obvious.  In fact, removing Higgs-boson couplings into QCD, one arrives at a scale invariant theory, which, classically, can't support any masses at all.  This contribution sketches forty years of developments in QCD, which suggest a solution to the puzzle, and highlight some of the experiments that can validate the picture.
}
%
\maketitle
\section{Introduction}
\label{intro}
The 2013 Nobel Prize in Physics was awarded to Englert and Higgs following discovery of the Higgs boson at the Large Hadron Collider (LHC) \cite{Englert:2014zpa, Higgs:2014aqa}.  With this discovery, the Standard Model of Particle Physics (SM) became complete; and some began to ask: ``Where do we go from here?''  The beginning of an answer lies in recognising just what the SM is.

As formulated, the SM offers a description of all known fundamental physics except gravity; and since gravity has no discernible effect when particles are studied a few at a time, one can largely neglect it when considering subatomic physics \cite{Politzer:2005kc}.  Since it discovery at the LHC \cite{Aad:2012tfa, Chatrchyan:2012xdj}, the Higgs Boson has been promoted to the \emph{Centre of Things} -- see, \emph{e.g}., Ref.\,\cite{SymmetryMagazine}.  In the lead image of that article it is apparent that the SM has 17 particles.  It also has 19 parameters, most of which relate to the Higgs and all of which must be determined through comparison with experiment.  So, although the SM supposedly describes the most powerful forces in Nature, with so many free parameters, it is somewhat unsatisfactory.

Many will have heard one or another variant of the oft made remark: ``All particles gain their mass from a fundamental field associated with the Higgs boson.''  There is some truth in this.  The Higgs does seem to be the source of the mass of elementary particles, \emph{e.g}., the electron; but it is responsible for $<2$\% of the mass of more complex things, like the proton.  The mass of the vast bulk of visible material in the Universe has a different source.

Strong interactions in the SM are expected to be explained by quantum chromodynamics (QCD) \cite{Marciano:1979wa}.  The Higgs reaches into QCD: it is the source of the Lagrangian current masses of the six known quarks.   These are 6 of the SM's 19 parameters.  However, only 2 of those 19 parameters are intrinsic to QCD.  One of them is $\theta_{\rm QCD}$, which appears to be (almost) zero because the nucleon electric dipole moment is, as yet, unmeasurably small \cite{Chupp:2017rkp}.  This leaves just one parameter in QCD that needs to be fixed -- a mass-scale, typically denoted $\Lambda_{\rm QCD}$.  So, perhaps science has a chance of understanding at least this one piece of the SM?

Or, perhaps not.  In many ways, QCD is the most important piece of the SM because it is supposed to describe all nuclear physics.  It is formulated in terms of matter fields -- six current quarks -- and gauge bosons -- eight gluons.  Yet, fifty years after the discovery of quarks \cite{Taylor:1991ew, Kendall:1991np, Friedman:1991nq, Friedman:1991ip}, science is only just beginning to understand how QCD moulds the basic elements of nuclei, \emph{viz}.\ pions, neutrons, protons, \emph{etc}.  Moreover, there are controversies, as theory begins to predict quantities that hitherto were only inferred from measurements via phenomenological fits.

To proceed, let's assume that quantum gauge field theories are the right paradigm for understanding Nature.  It is then important that we find ourselves in a Universe where time and space give us four (noncompact) dimensions.  $D=4$ is a critical point because quantum field theories in $D\neq 4$ dimensions possess an explicit mass-scale: their couplings carry a mass dimension.  Theories in $D>4$ are characterised by uncontrollable ultraviolet divergences.  On the other hand, those in $D<4$, while superconvergent, are afflicted with a hierarchy problem: dynamical effects typically contribute $\lesssim 10$\% to any mass-dimensioned quantity \cite{Appelquist:1981vg, Appelquist:1986fd, Bashir:2008fk, Bashir:2009fv, Braun:2014wja}.

In contrast, omitting Higgs couplings, the $D=4$ SM is built from scale-invariant classical field theories.  Each is renormalisable, a process that introduces a mass scale \cite[Ch.\,III]{Pascual:1984zb}.  The size of that mass is not determined by the theory.  So, what determines the natural scale for visible matter?  We know it is $m_{\rm Nature} \approx m_{\rm proton} \approx 1\,$GeV; but how much ``wiggle room'' is there in that number, \emph{i.e}., what is the window of values for $\delta m_{\rm Nature}$ such that $m_{\rm Nature}\to m_{\rm Nature}+\delta m_{\rm Nature}$ would still deliver a habitable Universe?  Perhaps an ultimate theory of Nature will answer these questions, but no one can be sure that such a theory exists.

This brings us to the central points.  Namely, the existence of our Universe depends critically on, \emph{inter alia}, the following empirical facts.
(\emph{a}) The proton is massive, by which one means that the mass-scale for strong interactions is much larger than that of electromagnetism.
(\emph{b}) It is also absolutely stable, despite being a composite object constituted from three light valence quarks.  (Here, light means that the proton's valence quarks have masses not much more than that of the electron \cite{Workman:2022ynf}.)
(\emph{c})
On the other hand, the $\pi$-meson is unnaturally light: its mass is commensurate with that of the $\mu$-lepton, despite the $\pi$ being a strongly interacting composite object built from a valence-quark and valence antiquark.
These things and their consequences are, supposedly, written in the QCD Lagrangian; but they are far from obvious.  Indeed, they are examples of emergence \cite{Roberts:2020udq, Roberts:2020hiw, Roberts:2021xnz, Roberts:2021nhw, Binosi:2022djx, Papavassiliou:2022wrb, Ding:2022ows}, \emph{viz}.\ low-level rules producing high-level phenomena, with enormous apparent complexity.

\section{Emergence of Hadron Mass}
\label{sec-1}
\begin{figure}[!t]
\hspace*{-1ex}\begin{tabular}{ll}
{\sf A} & {\sf B} \\[-2ex]
\includegraphics[clip, width=0.46\textwidth]{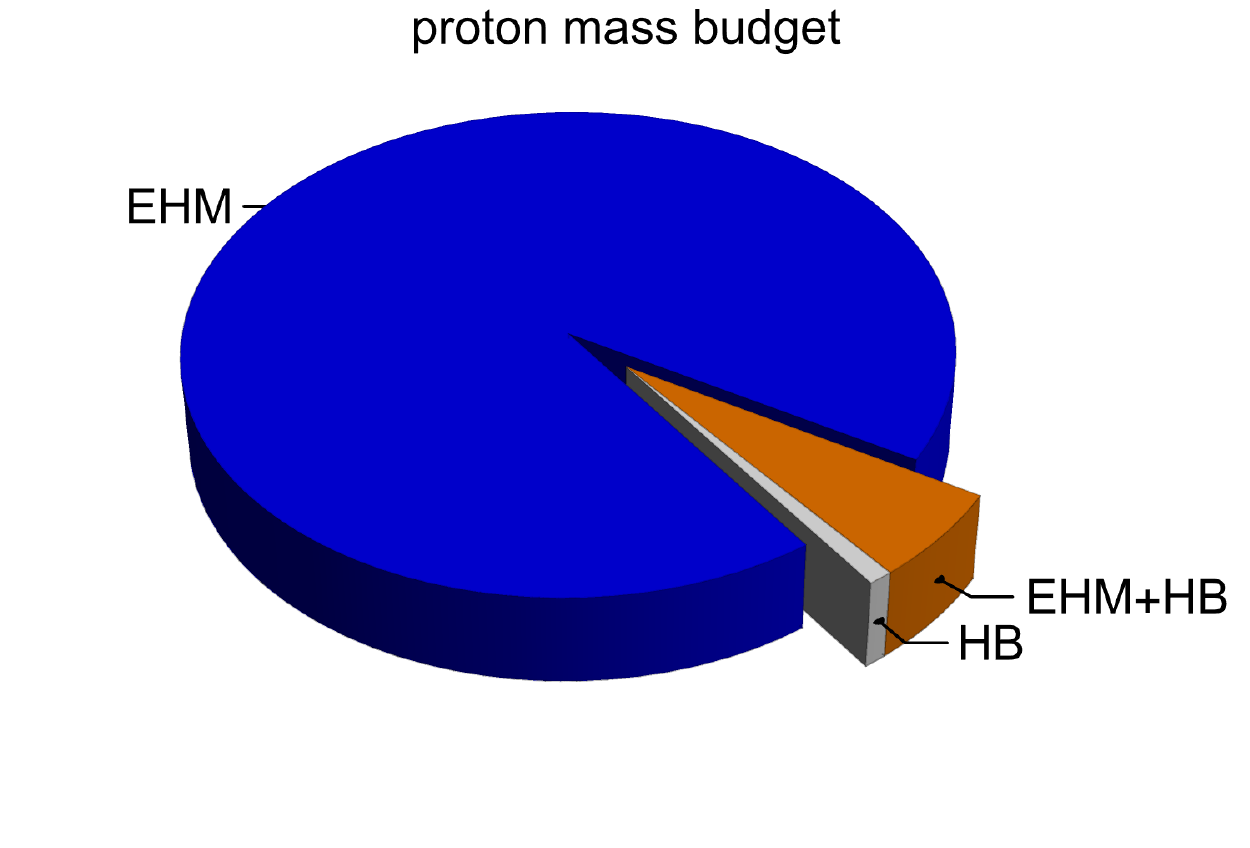} &
\includegraphics[clip, width=0.46\textwidth]{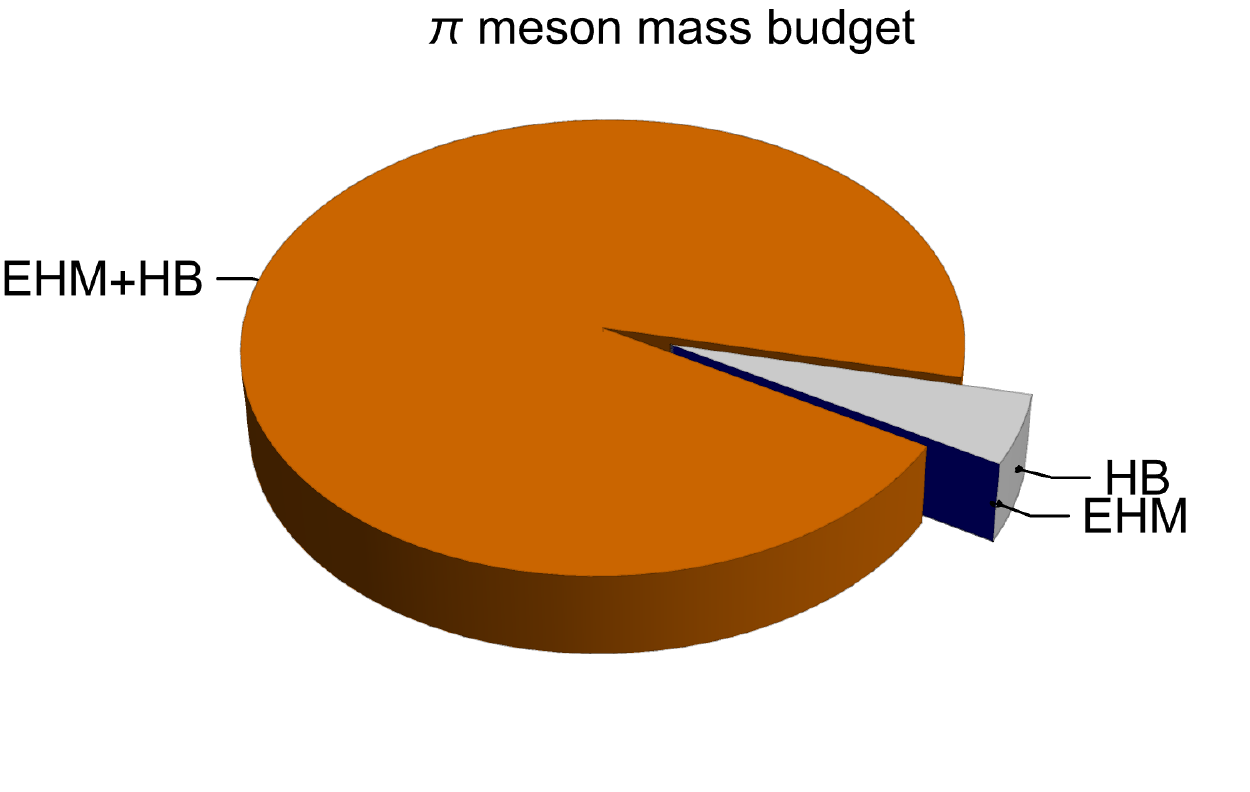}
\end{tabular}
\caption{\label{Fmassbudget}
Mass budgets:
\mbox{\sf A}\,--\,proton;
\mbox{\sf B}\,--\,pion.
Each is drawn using a Poincar\'e invariant decomposition, with the numerical values listed elsewhere \cite[Table~1]{Ding:2022ows}: HB -- contribution owing solely to the Higgs boson; EHM -- contribution from emergent hadron mass; EHM+HB -- mass generated by constructive interference between these two sources of mass.
}
\end{figure}

It is now plain that the SM has one known source of mass, \emph{i.e}., the Higgs boson (HB).  Its contributions are critical to the evolution of our Universe.  However, as shown in Fig.\,\ref{Fmassbudget}A, the Higgs is responsible for only 9\,MeV in the 939\,MeV of proton mass; hence, just 1\% of the mass of the visible material in the Universe.  Plainly, Nature has another, very effective mechanism for producing mass, which has come to be known as emergent hadron mass (EHM).  Alone, EHM is responsible for 94\% of $m_{\rm proton}$.  The remaining 5\% arises from constructive interference between these two sources.

At this point, many will be asking: ``But what is the origin of EHM?''  It's a key question; many others are related.  For instance: How is EHM connected with gluon and quark confinement -- part of the \$1-million prize Millennium Problem \cite{millennium:2006}; with dynamical chiral symmetry breaking -- the foundation stone for the existence of Nambu-Goldstone bosons; and what is the HB's role in modulating the observable properties of matter?  This last issue is critical because, \emph{e.g}., without the Higgs, all current-quarks would be massless and the $\pi$- and $K$-meson would be indistinguishable.  So take a look at the $\pi$-meson mass budget in Fig.\,\ref{Fmassbudget}B.  There is no EHM-only contribution in this case because, being a Nambu-Goldstone boson, the pion is massless in the chiral limit, \emph{i.e}., with HB couplings removed from QCD.  Even using realistic couplings, the HB part is still small -- just 5\%.  But EHM has a prodigious indirect impact: EHM+HB interference is responsible for 95\% of $m_\pi$ \cite[Sec.\,V]{Roberts:2020udq}.

In defining QCD, one begins with a one-line Lagrangian density:
\begin{equation}
\label{QCDdefine}
{\mathpzc L}_{\rm QCD}  = \sum_{{\mathpzc f}=u,d,s,\ldots}
\bar{q}_{\mathpzc f} [\gamma\cdot\partial
    + i g \tfrac{1}{2} \lambda^a\gamma\cdot A^a+ m_{\mathpzc f}] q_{\mathpzc f}
    + \tfrac{1}{4} G^a_{\mu\nu} G^a_{\mu\nu},\\
%
%
\end{equation}
where
$\{q_{\mathpzc f}\,|\,{\mathpzc f}=u,d,s,c,b,t\}$ are fields associated with the six known flavours of quarks; $\{m_{\mathpzc f}\}$ are their current-masses, generated by the Higgs boson;
$\{A_\mu^a\,|\,a=1,\ldots,8\}$ represent the gluon fields, whose matrix structure is encoded in $\{\tfrac{1}{2}\lambda^a\}$, the generators of SU$(3)$ in the fundamental representation;
$G^a_{\mu\nu} = \partial_\mu A^a_\nu + \partial_\nu A^a_\mu - g f^{abc}A^b_\mu A^c_\nu$ is the gluon field strength tensor;
and $g$ is the \emph{unique} QCD coupling, using which one conventionally defines $\alpha = g^2/[4\pi]$.
A peculiar feature of Eq.\,\eqref{QCDdefine} is that this Lagrangian density is expressed in terms of gluon and quark parton fields, which are \emph{not} the degrees-of-freedom measured in detectors.  This raises new questions.  Namely, what are the asymptotic, detectable degrees-of-freedom; and how are they built from the Lagrangian partons?  If such questions can be answered, then we might learn whether QCD is the theory of strong interactions; indeed, whether it is really a theory.  These things are crucial, because, hitherto, science has not delivered a mathematically well defined four dimensional quantum field theory.

The only mass scales manifest in Eq.\,\eqref{QCDdefine} are the current-quark masses, $\{m_{\mathpzc f}\}$.  Insofar as the proton is concerned, only up $(u)$ and down $(d)$ quarks are immediately relevant because the proton is a bound-state of the valence-quark combination $u+u+d$.  As apparent from Fig.\,\ref{Fmassbudget}, the current mass of any one of these quarks is just $1/300^{\rm th}$ of $m_{\rm Nature}$, \emph{viz}.\ more than two orders-of-magnitude smaller.  Plainly, the HB generated mass is very far removed from the natural scale for strongly-interacting matter.  Moreover, no amount of staring at ${\mathpzc L}_{\rm QCD}$ can reveal that scale.  This is in stark contrast to quantum electrodynamics (QED), in which, \emph{e.g}., the electron mass that appears explicitly in ${\mathpzc L}_{\rm QED}$ provides the mass unit for the spectrum of hydrogen energy levels.

So, let's remove the current-quark masses, \emph{i.e}., suppress HB couplings into QCD.  Then no explicit scales remain and the action associated with Eq.\,\eqref{QCDdefine} is scale invariant.  There is no dynamics in a scale invariant theory; only kinematics survive, and the theory appears the same at all length scales.  Thus, isolated clumps of material can't form, which means bound states are impossible.  Consequently, our Universe can't exist.  As noted above, the Higgs doesn't solve this problem because $u$, $d$ quarks are \emph{light} and the masses of protons and neutrons -- the kernels of all visible matter -- are $>100$-times larger than anything the Higgs produces in this quark sector.  Thus, the question of how our Universe came to be is inseparable from that of ``How do nucleons become so massive?''  (A detailed discussion of this and the following points in this section may be found in Ref.\,\cite{Roberts:2016vyn}.)

All theories have an energy-momentum tensor, $T_{\mu\nu}$ -- energy and momentum conservation are expressed in the divergence relation $\partial_\mu T_{\mu\nu}=0$.  The Noether current associated with spacetime dilations is ${\mathpzc D}_\mu = T_{\mu\nu} x_\nu$; and in a scale invariant theory, $\partial_\mu {\mathpzc D}_\mu = T_{\mu\mu} \equiv 0$.
In quantising chromodynamics, a mass scale is introduced by the process of regularisation and renormalisation of ultraviolet divergences.  This entails that the constants in ${\mathpzc L}_{\rm QCD}$, \emph{i.e}., coupling and current masses, come to depend on the renormalisation scale, hereafter denoted $\zeta$.  Let's switch off the HB couplings again, then one need only consider the QCD fine structure constant: $\alpha \to \alpha(\zeta)$.  In this case, some straightforward algebra involving Eq.\,\eqref{QCDdefine} reveals that, under the change of scale $\zeta \to {\rm e}^\sigma \zeta$, $\alpha \to \sigma \alpha \beta(\alpha)$, where $\beta(\alpha)$ is the QCD ``$\beta$-function'', and
\begin{equation}
\label{TraceAnomaly}
{\mathpzc L}_{\rm QCD} \to \sigma \, \alpha \beta(\alpha) \, \frac{\delta {\mathpzc L}_{\rm QCD}}{\delta \alpha}
\Rightarrow \partial_\mu {\mathpzc D}_\mu =
\frac{\delta {\mathpzc L}_{\rm QCD}}{\delta \sigma} = \alpha \beta(\alpha) \, \frac{{\mathpzc L}_{\rm QCD}}{\delta \alpha} = \beta(\alpha) \tfrac{1}{4} G_{\mu\nu}^a G_{\mu\nu}^a = T_{\mu\mu}=: \Theta_0\,.
\end{equation}
Evidently, quantisation of our four-dimensional theory has introduced a nonzero value for the trace of energy-momentum tensor.  This is the (in)famous QCD trace anomaly.

The existence of a trace anomaly means that a mass scale must exist in QCD.  The key issue is whether, or not, one can compute that scale and/or understand its magnitude.  Prediction of the scale's value is beyond the SM.  On the other hand, measurement is straightforward.  Consider a proton with four-momentum $P$, represented by a state vector $|p(P)\rangle$, then in a Poincar\'e invariant theory:
\begin{equation}
\langle p(P) | T_{\mu\nu} | p(P) \rangle = - P_\mu P_\nu
\Rightarrow \langle p(P) | T_{\mu\mu} | p(P) \rangle = - P_\mu P_\mu = m_{\rm proton}^2
= \langle p(P) | \Theta_0 | p(P) \rangle .
\end{equation}
Thus, absent HB couplings, the entirety of the proton mass owes to the trace anomaly.  Looking at Eq.\,\eqref{TraceAnomaly}, then from this perspective, $\Theta_0$ measures the in-proton strength of gluon self-interactions; so, $m_{\rm proton} \approx 1\,$GeV is somehow generated completely by $m=0$ gluon partons.

Let's consider the other mass budget drawn above, \emph{viz}.\ that for the $\pi$-meson in Fig.\,\ref{Fmassbudget}B.  In the absence of HB couplings, $m_\pi = 0$; hence,
\begin{equation}
\label{PionAnomaly}
\langle \pi(q) | T_{\mu\nu} | \pi(q) \rangle = - q_\mu q_\nu
\Rightarrow \langle \pi(q) | T_{\mu\mu} | \pi(q) \rangle = - q_\mu q_\mu = m_{\pi}^2 = 0
= \langle \pi(q) | \Theta_0 | \pi(q) \rangle .
\end{equation}
Does this mean that the scale anomaly vanishes trivially in the pion state; to wit, gluons contribute nothing to the pion mass?  Some have claimed this -- see Ref.\,\cite[Sec.\,2]{Aguilar:2019teb}; but given that all the proton's mass owes to $\Theta_0$, it seems unlikely.  Far more plausibly, ``$0$'' owes to cancellations between different operator contributions to the in-$\pi$ expectation value of $\Theta_0$.  In physics, of course, such precise cancellations are never accidental.  Typically, they only arise because some symmetry is dynamically broken -- see Eq.\,\eqref{pigtr} below.

It is now abundantly clear that one cannot address the question of the origin of $m_{\rm proton}$ without simultaneously explaining how the $\pi$-meson remains massless.  The natural visible-matter mass-scale must emerge coextensively with apparent preservation of scale invariance in related systems.  Moreover, the in-pion expectation value of $\Theta_0$ is always zero, irrespective of the size of $m_{\rm proton}$.
The challenge to modern physics is to elucidate the manifold empirical consequences of the mechanism responsible so that the SM can be validated.
(This does not overlook the fact that no finite number of measurements can validate a theory.  Instead, it asks for SM boundaries to be pushed in new directions within strong interaction physics with a view to testing whether $D=4$ quantum non-Abelian gauge field theories might be a viable paradigm for extending the SM.)

\section{Three Pillars of EHM}
The path to understanding EHM was laid sixty years ago, with the first demonstration that gauge bosons could dynamically acquire mass without destroying any of the essential features of gauge theories \cite{Schwinger:1962tn, Schwinger:1962tp}.  Forty years ago, it was realised that this  ``Schwinger mechanism'' of gauge boson mass generation could be active in QCD \cite{Cornwall:1981zr}.  Since then, the idea has been refined \cite{Aguilar:2008xm, Boucaud:2008ky, Binosi:2009qm, Boucaud:2011ug, Aguilar:2015bud}, so that a detailed picture is now emerging.  Continuum theory and numerical simulations of lattice-regularised QCD both agree that owing to gluon self-interactions, gluon partons are transformed into gluon quasiparticles, which are characterised by a momentum dependent mass function that is large at infrared momenta.  The associated renormalisation group invariant scale is \cite{Cui:2019dwv}:
\begin{equation}
\label{gluonmass}
m_0 = 0.43(1)\,{\rm GeV} \approx \tfrac{1}{2} m_{\rm Nature}.
\end{equation}
Here one truly sees \emph{mass emerge from nothing}: an interacting $D=4$ theory, expressed using massless gluon partons, dynamically generates massive dressed gluon fields.  This is the first pillar of EHM.  Now let's consider its consequences.

\begin{figure}[!t]
\centering
\sidecaption
\includegraphics[clip, width=0.5\textwidth]{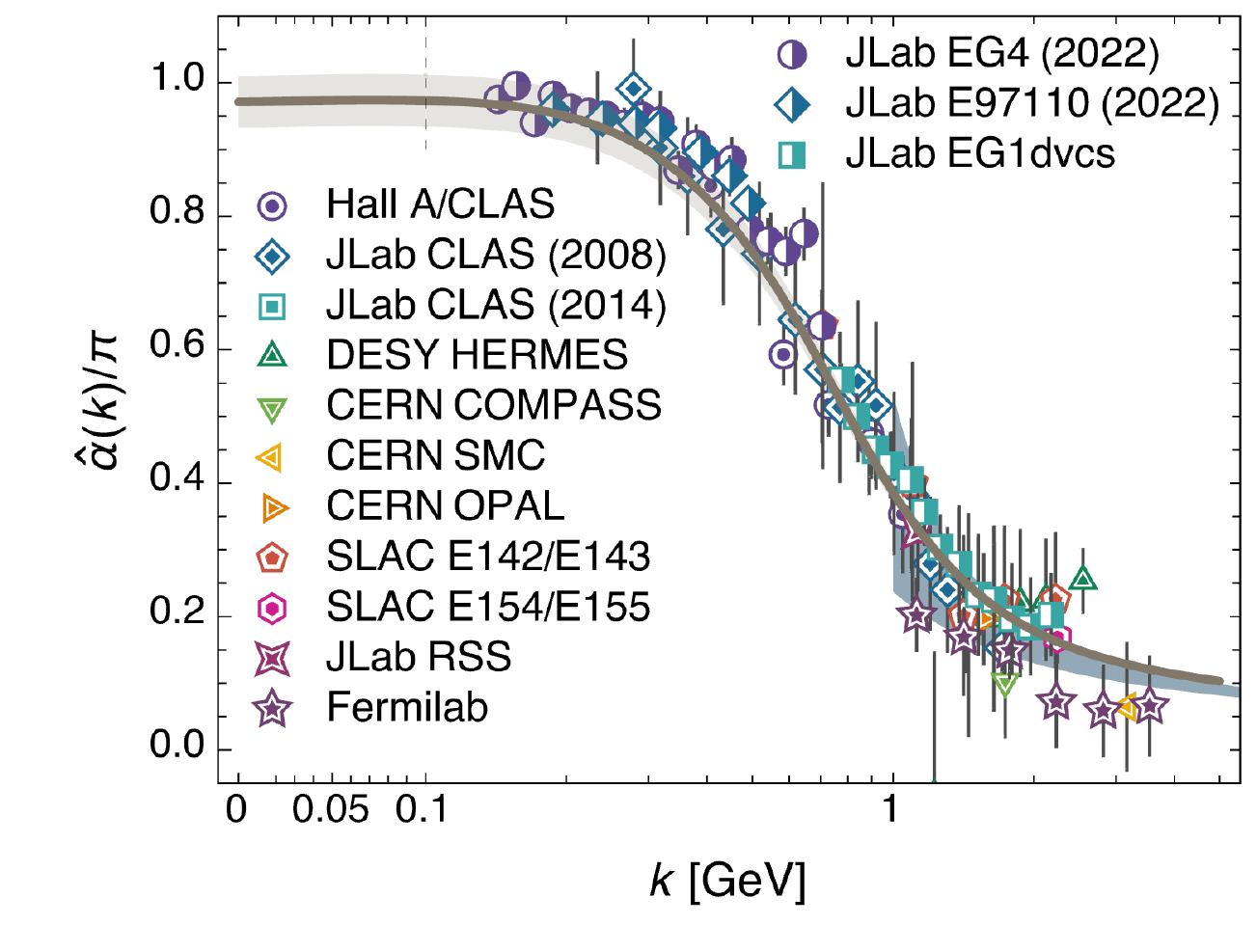}
\caption{\label{Falpha}
Process-independent effective charge, $\hat{\alpha}(k)/\pi$, obtained by combining modern results from continuum and lattice analyses of QCD's gauge sector \cite{Cui:2019dwv}.
Note the dramatic change in behaviour on $k \lesssim m_0$, as perturbative interactions are overwhelmed by EHM.
%
%
Existing data on the process-dependent charge, $\alpha_{g_1}$, are also drawn.  Their sources are discussed in Ref.\,\cite{Deur:2022msf} and the meaning of the comparison in Refs.\,\cite{Roberts:2021nhw, Ding:2022ows}.
(Image courtesy of D.\,Binosi.)}
\end{figure}

A crucial step on the road to presenting QCD as the theory of strong interactions was the discovery of asymptotic freedom \cite{Politzer:2005kc, Wilczek:2005az, Gross:2005kv}, \cite[Ch.\,7.1]{Pickering:1984tk}: the interaction between charge-carriers becomes weaker as the momentum-squared characterising the scattering process, $k^2$, becomes larger.  (QED exhibits the opposite behaviour \cite[Ch.~13]{IZ80}.)  The perturbative analysis of this coupling delivered a result that diverges at $k^2=\Lambda_{\rm QCD}^2< m_{\rm proton}^2$.  This is the ``Landau pole'', which can't be eliminated in perturbation theory.  On the other hand, modern methods in theory have enabled calculation of the unique QCD analogue of QED's Gell-Mann--Low effective charge, which extends the perturbative form onto the strong coupling domain.  The solution is drawn in Fig.\,\ref{Falpha}.  It is a parameter-free QCD prediction.  Physically, the generation of a gluon mass scale has ensured that long wavelength gluons are screened from interactions, eliminating the Landau pole, so that the QCD running coupling saturates to a finite value in the far infrared.  This running coupling is the second pillar of EHM.

The emergence of mass in QCD's gauge sector is communicated to the matter sector, with the appearance of a chiral-limit running mass for quarks, $M_0(k)$, that reaches a value at infrared momenta whose size is commensurate with the constituent-quark mass: $M_0(0)\lesssim m_0$.  Illustrated and discussed elsewhere \cite[Fig.\,2.5]{Roberts:2021nhw}, this is the third pillar of EHM.

These three pillars of EHM are expressed in every strong interaction observable, even if that is only by giving a new stability to perturbation theory through the emergence of dynamically generated infrared cutoffs.  The theory challenge is to elucidate their diverse measurable expressions; and the experimental challenge is to test the predictions and establish the soundness of the paradigm: simplicity to elegant complexity in the emergence of the most fundamental structures in Nature.

\section{EHM at Existing and Future Facilities}
To meet the theory and experimental challenges, science must move beyond the 100-year focus on proton structure.  Columbus did not discover the New World by meticulously charting the lagoon at \emph{Palos de la Frontera}; and QED was not revealed by careful studies of the hydrogen ground state alone.  Contemporary theory, which connects the three pillars of EHM to hadron observables, reveals extraordinary variations in structure between kindred and neighbouring states; so, science must use every tool at its disposal and develop new facilities that can expose the hand  of EHM in the spectrum of all hadrons and their structure.
The theory predictions and experimental opportunities are canvassed in numerous sources
\cite{Roberts:2020udq, Roberts:2020hiw, Roberts:2021xnz, Roberts:2021nhw, Binosi:2022djx, Papavassiliou:2022wrb, Ding:2022ows, Brodsky:2015aia, Aguilar:2019teb, Brodsky:2020vco, Chen:2020ijn, Anderle:2021wcy, Arrington:2021biu, Aoki:2021cqa, Quintans:2022utc}.  Herein, owing to limitations of space and time, I will only touch upon EHM as it is relates to future measurements of proton and pion structure functions.

Considering Eq.\,\eqref{PionAnomaly}, it is clear that the pion is special.  The fact that $\langle \pi(q) | \Theta_0 | \pi(q) \rangle \equiv 0$ in the absence of HB couplings is explained by a simple yet potent QCD identity; namely \cite{Maris:1997hd}
\begin{equation}
\label{pigtr}
\begin{array}{ccc}
\mbox{\rm two~body~problem} & & \mbox{\rm one~body~problem} \\
f_\pi^0 E_\pi^0(k;P) &=& B_0(k^2)
\end{array} \,.
\end{equation}
The left-hand side of this identity contains a product of terms that measures the leading component in the pion's bound-state amplitude; the right-hand side is the scalar piece of the dressed-quark self-energy; and this identity states that the pseudoscalar meson two-body problem is solved, almost completely, once the solution of the dressed-quark one body problem is known.  (The ``0'' super/subscript means HB couplings turned off.)  Eq.\,\eqref{pigtr} is a corollary of EHM: it is a sufficient and necessary condition for dynamical chiral symmetry breaking; provides the QCD foundation for the success of chiral effective field theory; and means that the properties of the (nearly) massless pion are the cleanest expressions of EHM in Nature.

\begin{figure}[!t]
\centering
\sidecaption
\includegraphics[clip, width=0.55\textwidth]{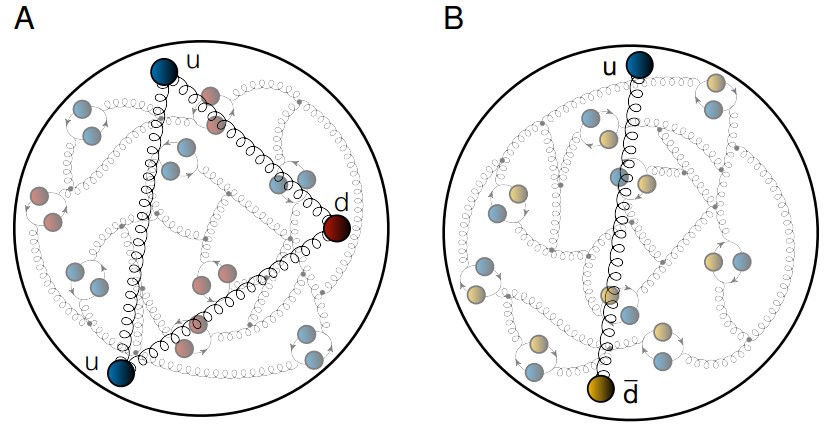}
\caption{\label{Fprotonpion}
{\sf A}. In terms of QCD's Lagrangian quanta, the proton, $p$, contains two valence up ($u$) quarks and one valence down ($d$) quark; and also infinitely many gluons and sea quarks, drawn here as ``springs'' and closed loops, respectively.
{\sf B}. The pion, $\pi^+$, contains one valence $u$-quark, one valence $\bar d$-quark, and, akin to the proton, infinitely many gluons and sea quarks.}
\end{figure}

Viewed from a perspective defined by ${\mathpzc L}_{\rm QCD}$ in Eq.\,\ref{QCDdefine}, protons and pions appear as sketched schematically in Fig.\,\ref{Fprotonpion}.  This structure is characterised by light-front momentum-fraction distribution functions (DFs) of valence quark partons, sea quark partons, and gluon partons, which can, in principle, be extracted from analyses of structure function measurements.  However, despite enormous expense of time and effort, much must still be learnt before the proton and pion may be considered understood in terms of these DFs.  Most simply, what are the differences, if any, between the distributions of partons within the proton and the pion?  Owing to a growing appreciation of the need to understand EHM, this question now has particular resonance as science attempts to determine how the obvious macroscopic differences between protons and pions are expressed in their DFs.

Regarding measurements that do not involve beam or target polarisation, it is known that there is a scale -- the hadron scale, $\zeta_{\cal H}<m_{\rm proton}$, at which valence-quark DFs in the proton and pion behave as follows \cite{Brodsky:1994kg, Yuan:2003fs, Cui:2021mom, Cui:2022bxn, Chang:2022jri}:
\begin{equation}
\label{LargeX}
{\mathpzc d}^p(x;\zeta_{\cal H}), {\mathpzc u}^p(x;\zeta_{\cal H}) \stackrel{x\simeq 1}{\propto} (1-x)^{\beta_{\cal H}^p=3}\,,
\quad
\bar {\mathpzc d}^\pi(x;\zeta_{\cal H}), {\mathpzc u}^\pi(x;\zeta_{\cal H})  \stackrel{x\simeq 1}{\propto} (1-x)^{\beta_{\cal H}^\pi=2}\,.
\end{equation}
The light-front momentum fraction is bounded: $0\leq x \leq 1$.  It subsequently follows from the QCD evolution equations \cite[DGLAP]{Dokshitzer:1977sg, Gribov:1971zn, Lipatov:1974qm, Altarelli:1977zs} that the large-$x$ power on the related glue DF is $\approx \beta_{\cal H}^{p,\pi}+1$ and that for sea quark DFs is $\approx \beta_{\cal H}^{p,\pi}+2$.  Moreover, as the scale increases to $\zeta > \zeta_{\cal H}$, these exponents all grow logarithmically.  However, such QCD constraints are typically ignored in fits to the world's data on structure functions and kindred observables \cite{Ball:2016spl, Hou:2019efy, Bailey:2020ooq, Novikov:2020snp, Barry:2021osv}.  This is fueling controversy, leading some to question the veracity of QCD \cite{Aicher:2010cb, Cui:2021mom, Cui:2022bxn}.
In addition, largely because pion data are scarce \cite[Table~9.5]{Roberts:2021nhw}, proton and pion data have never been considered simultaneously.

\begin{figure}[!t]
\hspace*{-1ex}\begin{tabular}{ll}
{\sf A} & {\sf B} \\[-2ex]
\includegraphics[clip, width=0.46\textwidth]{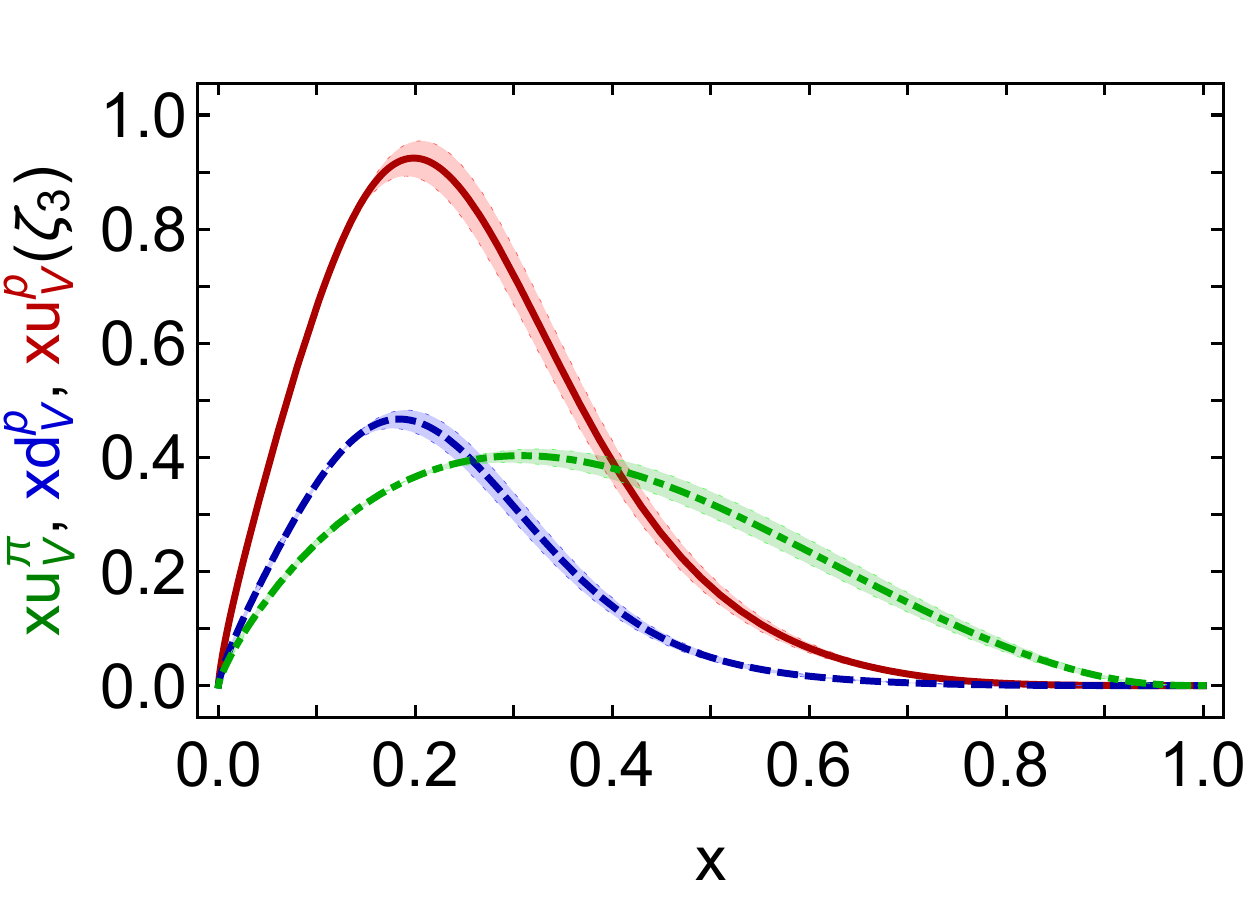} &
\includegraphics[clip, width=0.46\textwidth]{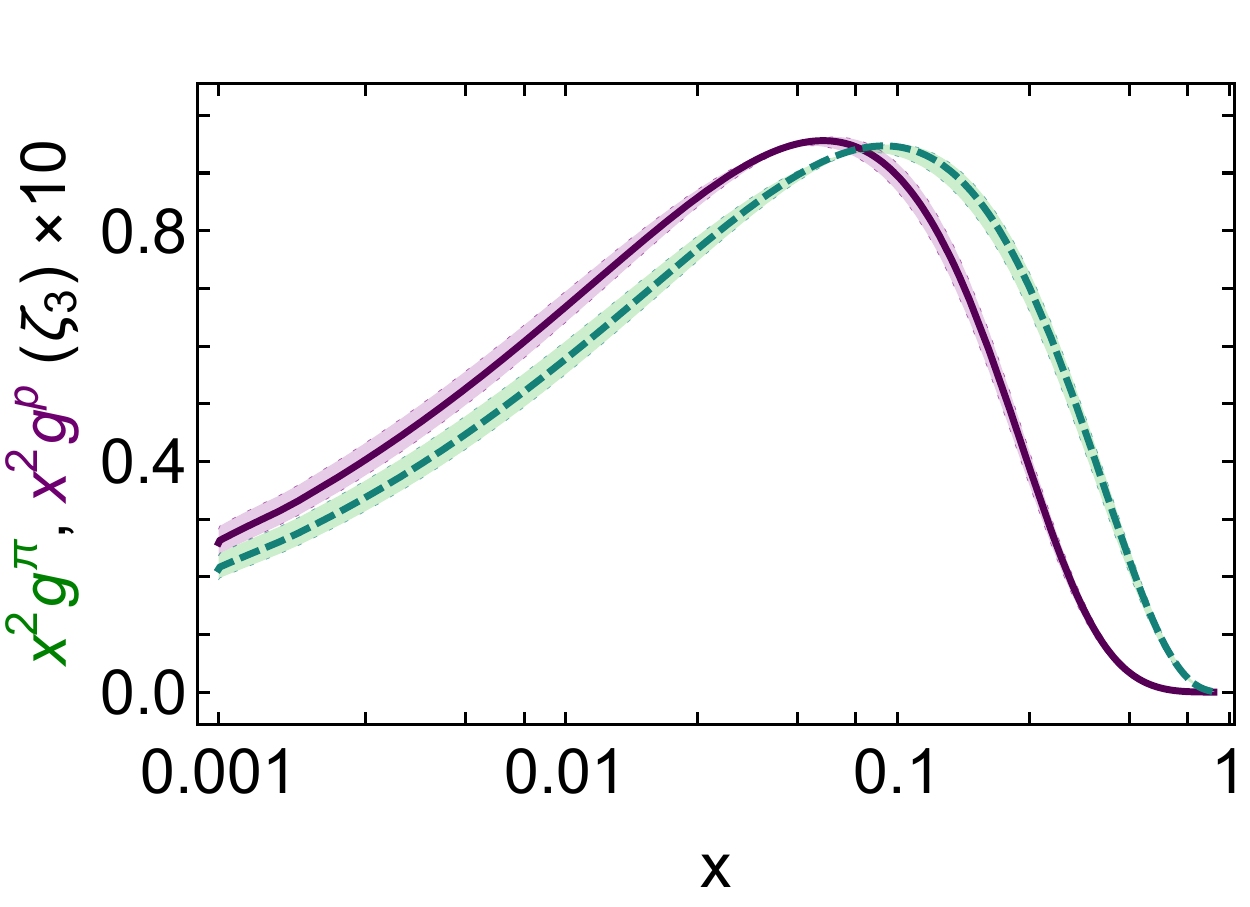}
\end{tabular}
\caption{\label{FDFs}
Distribution functions.
\mbox{\sf A}\,--\,valence quark DFs in the proton ($u$ quark, solid red; $d$-quark, dashed blue) and pion (dashed green);
\mbox{\sf B}\,--\,glue DFs in the proton (solid purple) and pion (dashed green).
In each case, the like coloured band expresses the response to a $\pm 5$\% variation in $\zeta_{\cal H}$.
}
\end{figure}

In this context, the unified array of results in Ref.\,\cite{Lu:2022cjx}, which uses a symmetry-preserving continuum Schwinger function method (CSM) to simultaneously predict the pointwise behaviour of all proton and pion DFs -- valence, glue, and four-flavour-separated sea -- is of significant value.  Fig.\,\ref{FDFs} displays predictions for valence-quark and glue DFs at the scale $\zeta=m_{J/\psi} = 3.1\,$GeV.  The following features are worth highlighting.
(\emph{i})
At $\zeta = \zeta_{\cal H}$, valence quark degrees-of-freedom carry all the hadron's momentum:
$\langle x \rangle_{{\mathpzc u}_p}^{\zeta_{\cal H}}=0.69$;
$\langle x \rangle_{{\mathpzc d}_p}^{\zeta_{\cal H}}=0.31$;
$\langle x \rangle_{{\mathpzc d}_\pi}^{\zeta_{\cal H}}=\langle x \rangle_{{\mathpzc u}_\pi}^{\zeta_{\cal H}}=0.5$.
This is the defining characteristic of $\zeta_{\cal H}$.
(\emph{ii})
Quark+quark (diquark) correlations within the proton, whose formation is driven by EHM, ensure ${\mathpzc u}^p(x;\zeta) \neq 2{\mathpzc d}^p(x;\zeta)$.  The predicted result for the ratio of these two functions agrees with modern experiments \cite{Segarra:2019gbp, Abrams:2021xum, Cui:2021gzg}.  Indeed, the data reveal that the proton must contain both scalar and axialvector diquarks: the probability that the data can be explained by a scalar-diquark-only proton wave function is $1/141,000$ \cite[Fig.\,27]{Ding:2022ows}.
(\emph{iii})
Proton and pion DFs have markedly different behaviour on $x>0.1$, which is called the valence-quark domain, with all pion DFs being markedly more dilated.  In fact, pion DFs are the most dilated in Nature.  This is a dramatic difference between proton and pion DFs.  It follows from Eq.\,\eqref{pigtr} and is a ``smoking gun'' signal for EHM.

It should also be stressed that the CSM predictions comply with QCD constraints on DF endpoint (low- and high-$x$) scaling behaviour, \emph{e.g}., Eq.\,\eqref{LargeX}.  This is in stark contrast to existing global fits to data, which ignore them; so, fail to deliver realistic DFs, even from abundant proton data, and return controversial fits to the limited pion data.  One may reasonably argue that only after imposing QCD constraints on future phenomenological data fits will it be possible to draw reliable pictures of hadron structure.  This is especially important for attempts to empirically expose structural differences between Nambu-Goldstone bosons and seemingly less complex hadrons.

EHM is expressed in numerous other observables.  For example, theory predicts that many excited states of the proton also contain pseudoscalar and vector diquark correlations.  This can be tested in resonance electroexcitation measurements \cite{Carman:2020qmb, Mokeev:2021dab, Mokeev:2022xfo}.  In addition, electroweak transitions between heavy+light mesons, in which HB couplings are the dominant source of mass, and light (lighter) final states, whose mass is far more reliant on EHM, provide ideal opportunities to measure the impact of interference between Nature's two mass-generating mechanisms \cite[Sec.\,10]{Ding:2022ows}.
In all these areas, progress demands a synergy between experiment, phenomenology, and theory.

\section{Conclusion}
QCD is an exceptional fundamental theory of natural phenomena.  Its Lagrangian degrees-of-freedom are not directly observable.  The Lagrangian's massless gluon partons become massive without ``human intervention''; and their mass guarantees the existence of a unique running coupling that is everywhere finite, ensuring a stable infrared completion of the theory.  Even in the absence of Higgs-boson coupling, the massless quarks become massive and interact to produce both massive hadrons and massless Nambu-Goldstone bosons.  These effects form the three pillars of emergence in QCD.  They are expressed in every strong interaction observable and can also be revealed via studies that contrast systems in which HB couplings provide the largest source of mass and those in which emergent mass is dominant.

Today, science is capable of building facilities that can validate these concepts, and thereby, possibly, elevate QCD, from its place as just one more effective field theory in the Standard Model, to a new position, as the first well-defined four-dimensional quantum field theory we've ever contemplated.  That step may open doors that lead far beyond the Standard Model because, while we have theories of many things, we are still asking whether there is a theory of everything.

\medskip

%
\noindent{\bf Acknowledgments}. This contribution is based on insights developed through collaborations with many people, to all of whom I am greatly indebted.
Work supported by:
National Natural Science Foundation of China (grant no.\,12135007).

%
%
%
%
%

\end{document}